\documentclass[sn-mathphys]{sn-jnl}%

\usepackage{grffile} %
\usepackage[numbers]{natbib}

\usepackage{enumitem,amssymb}
\newlist{todolist}{itemize}{2}
\setlist[todolist]{label=$\square$}
\usepackage{pifont}
\usepackage{booktabs}       %
\usepackage{xcolor}         %
\usepackage{listings}
\usepackage{overpic}
\usepackage{caption}
\usepackage{subcaption}
\usepackage[normalem]{ulem}
\usepackage[version=4]{mhchem}
\usepackage{amsmath}

\jyear{2022}%

\theoremstyle{thmstyleone}%

\theoremstyle{thmstyletwo}%

\theoremstyle{thmstylethree}%

\usepackage{bm}
\let\bmm\bm
\usepackage{xspace}		%
\usepackage{amssymb}	%
\usepackage{upgreek}
\usepackage{siunitx}

\newcommand{\vectornorm}[1]{\left\lVert#1\right\rVert}
\newcommand{\abs}[1]{\left\lvert#1\right\rvert}

\renewcommand{\lim}[1]{\underset{#1}{\operatorname{lim}}\;}
\makeatletter
 \begingroup
 \catcode`\_=\active
 \protected\gdef_{\@ifnextchar|\subtextup\sb}
 \endgroup
 \def\subtextup|#1|{\sb{\textup{#1}}}
 \AtBeginDocument{\catcode`\_=12 \mathcode`\_=32768 }
\makeatother

\newcommand{\bphi}{\ensuremath{\bmm{\upphi}}\xspace}

\newcommand{\ba}{\ensuremath{\mathbf{a}}\xspace}

\newcommand{\bG}{\ensuremath{\mathbf{G}}\xspace}

\renewcommand{\bm}{\ensuremath{\mathbf{m}}\xspace}

\newcommand{\bp}{\ensuremath{\mathbf{p}}\xspace}

\newcommand{\brr}{\ensuremath{\mathbf{r}}\xspace}

\newcommand{\bss}{\ensuremath{\mathbf{s}}\xspace}

\newcommand{\bU}{\ensuremath{\mathbf{U}}\xspace}
\newcommand{\bv}{\ensuremath{\mathbf{v}}\xspace}
\newcommand{\bV}{\ensuremath{\mathbf{V}}\xspace}

\newcommand{\bW}{\ensuremath{\mathbf{W}}\xspace}
\newcommand{\bx}{\ensuremath{\mathbf{x}}\xspace}

\usepackage[draft]{fixme}
\newcommand{\figref}[1]{\figurename~\ref{#1}}

\usepackage{letltxmacro}
\LetLtxMacro{\originaleqref}{\eqref}
\renewcommand{\eqref}{Equation~\originaleqref}

\begin{document}

\title[DeepDFT]{Equivariant graph neural networks for fast electron density estimation of molecules, liquids, and solids}

\author[1]{\fnm{Peter Bjørn} \sur{Jørgensen}}\email{pbjo@dtu.dk}

\author*[1]{\fnm{Arghya} \sur{Bhowmik}}\email{arbh@dtu.dk}

\affil*[1]{\orgdiv{Department of Energy Conversion and Storage}, \orgname{Technical University of Denmark}, \orgaddress{\street{Anker Engelundsvej 1}, \city{Kgs. Lyngby}, \postcode{2800}, \country{Denmark}}}

\abstract{%
Electron density $\rho(\vec{\brr})$ is the fundamental variable in the calculation of ground state energy with density functional theory (DFT). Beyond total energy, features and changes in $\rho(\vec{\brr})$ distributions are often used to capture critical physicochemical phenomena in functional materials. We present a machine learning framework for the prediction of $\rho(\vec{\brr})$. The model is based on equivariant graph neural networks and the electron density is predicted at special query point vertices that are part of the message passing graph, but only receive messages. The model is tested across multiple data sets of molecules (QM9), liquid ethylene carbonate electrolyte (EC) and LixNiyMnzCo(1-y-z)O2 lithium ion battery cathodes (NMC). For QM9 molecules, the accuracy of the proposed model exceeds typical variability in $\rho(\vec{\brr})$ obtained from DFT done with different exchange-correlation functionals.  The accuracy on all three datasets is beyond state of the art and the computation time is orders of magnitude faster than DFT.
}


\maketitle
\section{Introduction}

Simulations are as critical as experiments now in materials discovery. At the atomic scale, quantum mechanics based simulations are frequently used in the computational search of novel functional materials and molecules\cite{marzari2021electronic}. Within the well-known cost-accuracy trade-off associated with such methods, Kohn-Sham density functional theory (DFT) is the most widely used method due to the right balance between computational cost and accuracy. The electron density $\rho(\vec{\brr})$ is one of the fundamental variables in the state of the art iterative scheme of DFT. The electron density uniquely determines the ground state properties of a system\cite{payne1992iterative}. DFT is an $O(n^3)$ complexity method and thus is limited to a few hundred atoms in system size that can be simulated. System size limits prohibit us from fully exploiting DFT for simulating critical technologically and scientifically important systems. For example, one would need large simulation cells for portraying engineering ceramics (with many types of atoms in small fractions) or mixed liquid electrolytes (with many component molecules and additives). Not just system size, in many materials design problems, the enormity of phase space to be explored can also be a bottleneck in using DFT. 
Total energy is the most commonly used output from DFT simulations. Significant recent developments towards high accuracy machine learning potentials for molecules and condensed matter phases have been able to provide QM accuracy total energy at a much lower computational cost\cite{unke2021machine,deringer2021gaussian,behler2021four}. However, for functional materials, electronic structure is important as well. Electronic density distribution and its modulation due to structural and chemical modifications are descriptors for many chemical properties. For example, Bader charge analysis is frequently used to understand redox processes and related phenomena in intercalation battery cathodes\cite{kondrakov2017charge, dixit2017origin, varanasi2014tuning, kuo2020origin}. Charge densities are critical for solar cell materials properties as well\cite{chang2021lead,barragan2017atomic}. Similarly, charge density redistribution is often used to understand trends in catalytic activity\cite{castellani2009density, palmer2019methane, zaffran2016metal, vasileff2019selectivity}. In liquid electrolytes, the analysis of gradients and other features in charge density helps us understand intermolecular interactions\cite{cao2017reversibility, mangiatordi2011modeling, del2014electron, armakovic2015dft}. Charge density also gives us access to functional properties (through surrogate models) that are costly to calculate directly. For example, charge density maps can give us direct access to optimum intercalation sites\cite{shen2020charge} as well as ion migration pathway and barriers\cite{kahle2020high}. We exemplify a few of the many possible ways fast machine learning prediction of charge density can help us find new better functional materials. We can explore larger phase space as well as evaluate functional properties of materials that require large simulation boxes.

Electron density is inherently more information rich than total energy and therefore learning from the density could lead to machine learning models that generalize better from small datasets.
For example, \cite{tsubakiQuantumDeepField2020} and \cite{lewisLearningElectronDensities2021} both found that learning the electron density and then predicting the total energy gives better accuracy when extrapolating from small to large systems in comparison to direct energy prediction.
In the last few years, a number of articles have been published on electron density prediction. Pioneering works by \cite{brockherdeBypassingKohnShamEquations2017, bogojeski2018efficient} use a basis representation for the density and predict the basis function coefficients using kernel ridge regression. The model's efficacy was demonstrated on molecular dynamics trajectories of small molecules, but by construction the model is not transferable to new molecules.
A transferable model based on symmetry-adapted Gaussian process regression\cite{sa-gpr} (SA-GPR) was introduced later \cite{grisafiTransferableMachineLearningModel2019,fabrizioElectronDensityLearning2019}.
The transferability is achieved by decomposing the density into atom-centered contributions and the local environment around each atom is mapped to a set of basis coefficients using the SA-GPR framework.
One of the downsides of kernel regression is that the computational complexity of the model grows cubically with the number of training examples and in most practical problems we will need thousands of training examples to cover the system of interest.

Deep neural network models are highly flexible and are generally well suited for absorbing large datasets.
A 3D convolutional neural network has been \cite{sinitskiy2018deep} trained with thousands of small molecules.
However, by using a voxel based 3D U-Net \cite{RFB15a} architecture the model is dependent on the image resolution and is not equivariant to rotations.
Equivariance to rotation has been achieved in different ways. The aforementioned SA-GPR model \cite{grisafiTransferableMachineLearningModel2019, fabrizioElectronDensityLearning2019} has symmetry built into its kernel function.

For deep learning models, equivariance has been achieved by constructing a fingerprint for every point in space that is invariant to rotations around the point for which the fingerprint has been created, but not to rotations around other points.

\cite{zepeda-nunezDeepDensityCircumventing2019, chandrasekaranSolvingElectronicStructure2019, kamalChargeDensityPrediction2020, ALRED20183}.
The models are local, which means that a cutoff distance defines the range for which the atoms no longer influence the electron density.
Message-passing neural networks \cite{gilmerNeuralMessagePassing2017, schuttSchNetDeepLearning2018} provide a mechanism for propagating atomic interactions over longer distances in a computationally efficient manner, by computing local messages that represents an atom and its environment and then propagate this information via the edges of a graph representation of the system.
Message-passing neural networks are also being applied to the electron density prediction problem.
In the work by \cite{gongPredictingChargeDensity2019} a message passing network is used as part of the algorithm, but a new graph, representing the local neighborhood, is created for every point in space, which makes the method computationally inefficient and the model was therefore only trained on a relatively small number of points.
A more efficient approach was later presented by 
\cite{cuevas-zuviriaAnalyticalModelElectron2020}\cite{cuevas-zuviriaMachineLearningAnalytical2021} in which atom-centered density contributions are predicted with a message passing neural network.
This approach allows the energy of the system to be calculated analytically, but the accuracy of the density predictions are inferior to our previously published framework, DeepDFT, which predicts the densities directly point by point \cite{jorgensen2020deepdft}.
In the previous message passing solutions \cite{gongPredictingChargeDensity2019, cuevas-zuviriaAnalyticalModelElectron2020, cuevas-zuviriaMachineLearningAnalytical2021, jorgensen2020deepdft} an invariant representation is used.
However, in this first generation of message passing algorithms, the models are unable to resolve angular information through the message passing \cite{pmlr-v139-schutt21a}.
Recent developments in equivariant message passing neural networks \cite{cohenGroupEquivariantConvolutional, kondorClebschGordanNets2018, thomas2018tensor, NEURIPS2019_03573b32, pmlr-v139-schutt21a, qiaoUNiTEUnitaryNbody2021} make it possible to propagate directional information through the message passing steps from which angular information can be extracted.
Common for the equivariant message passing models is that the hidden states of the graph nodes are now representing directional vectors (in three dimensions) that rotate with the rotation of the molecule.

In this work, we present the equivariant DeepDFT model, which is a machine learning model for predicting the electron density $\rho(\vec{\brr})$.
The model is based on equivariant message passing on a graph and uses special probe nodes inserted into the graph, for which the density is computed.
In contrast to the OrbNet-Equi model \cite{qiaoUNiTEUnitaryNbody2021},
which uses features calculated by the GFN1-xTB semiempirical electronic structure method, our method is purely data driven in the sense that the only inputs required to make a prediction are the atomic numbers and the coordinates of the atoms (including the unit cell parameters for periodic structures).
We also do not induce any bias in terms of using a predefined basis set for the density, i.e., the density is purely learned from data examples.

Previous models have shown the ability to predict electron density on a number of different systems, including molecular dynamics of a single molecule or slab \cite{brockherdeBypassingKohnShamEquations2017, chandrasekaranSolvingElectronicStructure2019}, different hydrocarbon molecules \cite{grisafiTransferableMachineLearningModel2019, kamalChargeDensityPrediction2020}, large datasets of small organic molecules \cite{sinitskiy2018deep, cuevas-zuviriaAnalyticalModelElectron2020, qiaoUNiTEUnitaryNbody2021}, carbon nanotubes \cite{ALRED20183}, crystalline polymers and zeolites \cite{gongPredictingChargeDensity2019} and peptides \cite{fabrizioElectronDensityLearning2019, cuevas-zuviriaMachineLearningAnalytical2021}.
To showcase the universal applicability, we benchmark the equivariant DeepDFT model on three diverse datasets, (a) the QM9 dataset \cite{Ruddigkeit2012-dc, Ramakrishnan2014-ey} often used for benchmarking molecular machine learning models, which is a large dataset of 134k small organic molecules (b) a dataset of mixed transition metal layered oxide lithium ion battery cathode materials and (c) a dataset consisting of a molecular dynamics trajectory with ethylene carbonate molecules - a commonly used organic electrolyte.

\section{Results}
\subsection{Invariant and equivariant message passing models}
\begin{figure}[htbp]
	\centering\includegraphics[width=0.8\textwidth]{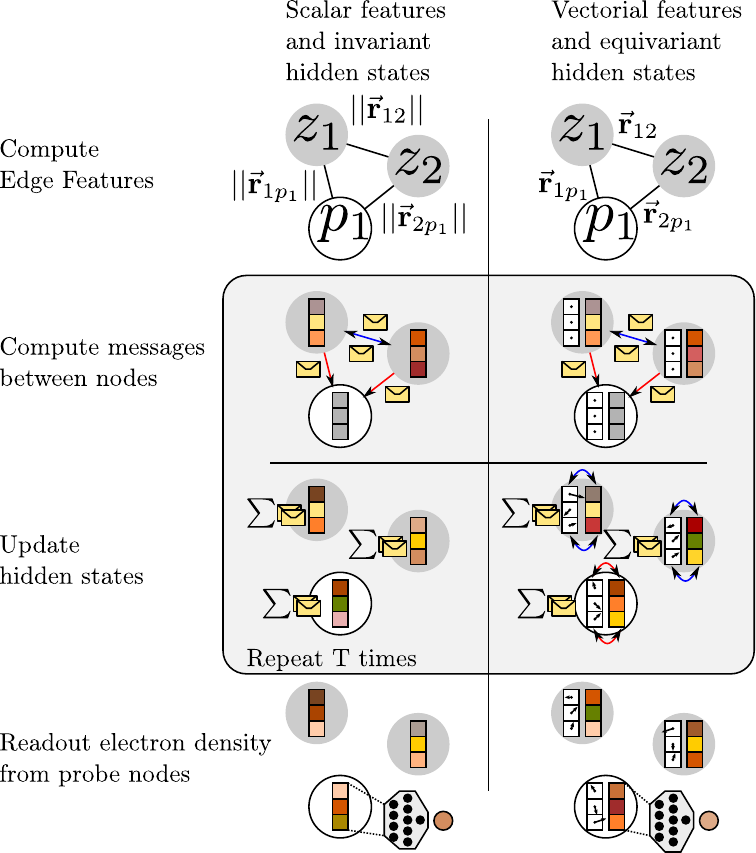}
	\caption{Conceptual overview of the two message passing architectures used in DeepDFT. The left column illustrates the invariant DeepDFT model and the right column illustrates the equivariant DeepDFT model.}
	\label{fig:msgpassing_model_equivariant}
\end{figure}

We have developed and trained two models for charge density prediction with different architectures, but both work on the principles of message passing on molecular graphs\cite{gilmerNeuralMessagePassing2017}. In this section we will be referring to both of the models - the invariant DeepDFT model and the equivariant DeepDFT model.
A conceptual overview of the equivariant and invariant DeepDFT models are shown in \figref{fig:msgpassing_model_equivariant}.

The models use a 3D-embedded graph representation of the molecule or crystal structure.
The graph has a vertex for each atom in the molecule or for each atom in the crystal structure unit cell.
Edges are defined by a constant cutoff distance, i.e. we draw an edge between vertices if the distance between them is less than a certain cutoff distance (chosen here to be \SI{4}{\angstrom}).
The edges may cross the periodic boundary as in quotient graphs \cite{chungNomenclatureGenerationThreeperiodic1984, kleeCrystallographicNetsTheir2004}.
Special probe vertices, that only accept incoming edges, are placed at each query point where electron density prediction is to be made.
Each vertex has a hidden state, which represents the atom or probe and its environment and the state is initialized based on the atom type or is zero for the probe vertices.
The vertices interact by receiving messages from other vertices via the incoming edges. After each exchange of messages, the vertices update their hidden state based on the sum of incoming messages.
Artificial neural networks (ANN) are used to model - (a) how the content of the messages depend on sending/receiving vertices and (b) how the hidden state updates depend on the message sum. ANNs representing these interactions can be trained using data examples.
After a number of interaction steps $T$ another neural network is used to map the hidden state of each probe vertex to the predicted density at that point in space.
In the invariant version of the DeepDFT model, the edge feature is the distance between the vertices, while in the equivariant version, the distance as well as the direction of the edges are used as features.
The directional features affect the hidden state of the vertices. To maintain the directionality of the hidden states, the equivariant model contains two sets of hidden states, an array of vector valued features for each vertex in addition to the array of scalar valued features for each vertex.
The equivariance of the hidden states is preserved by restricting the allowed operations to scaling, inner products with other equivariant features and the addition of linear combinations of other equivariant features.
The details of the invariant DeepDFT model are described in our previous NeurIPS workshop paper \cite{jorgensen2020deepdft} and more details of the equivariant model can be found in the methods section.

\subsection{Dataset and setup}
To assess the models, we use three diverse datasets.
The first is the QM9 dataset \cite{Ruddigkeit2012-dc, Ramakrishnan2014-ey} (134k small molecules with up to nine heavy atoms (CNOF)) that is widely used for benchmarking machine learning models for molecular property prediction.
Additionally, we also train and test with electron density data from crystalline and liquid state materials.
1:~A class of industrially important mixed transition metal layered oxide lithium ion battery (LIB) cathode materials.
Configurations are generated through random crystal site occupation of transition metal ions (Ni/Mn/Co) and lithium/vacancy to represent varieties of chemistry and lithiation states.
2:~Liquid ethylene carbonate - the most used LIB electrolyte.
12000 disordered configurations are generated through high temperature (3000K) accelerated molecular identity preserving molecular dynamics.
In all three cases, the electron densities are obtained using the VASP code \cite{hafner2008ab}. See the methods section for more details on the computational setup.

\subsection{Prediction accuracy}
In this numerical experiment we assess the average prediction accuracy of the model using the three aforementioned datasets. We split the datasets into training, validation (for model selection) and test set with uniform random splitting. The sizes of the splits are shown in table \ref{tab:datasets}.
To evaluate the model's accuracy in predicting charge density for a given atomic structure, we integrate the mean absolute error (MAE) over the whole simulation box normalized by the total number of electrons (\eqref{eq:mae}) following recently published work\cite{grisafiTransferableMachineLearningModel2019, fabrizioElectronDensityLearning2019}. %
To obtain the metric for a given test VASP electron density, the model is probed at every point corresponding to the positions of the electron density grid saved by VASP, and the integrals are evaluated numerically as a sum
. 
The DFT calculated density is used as ground truth $\rho(\vec{\brr})$ and the trained models output the predicted density $\hat{\rho}(\vec{\brr})$.
\begin{equation}
	\varepsilon_|mae| = \frac{\int_{\vec{\brr} \in V} \abs{ \rho (\vec{\brr}) - \hat{\rho} (\vec{\brr}) }}{\int_{\vec{\brr} \in V} \abs{\rho(\vec{\brr}) }}
	\label{eq:mae}
\end{equation}
The MAE of the models are shown in Table~\ref{tab:datasets}. 
As a baseline model we use the superposition as of atomic densities, which is used by VASP to initialize the electron densities.

\begin{figure}[tp]
	\centering\includegraphics[width=1.0\textwidth]{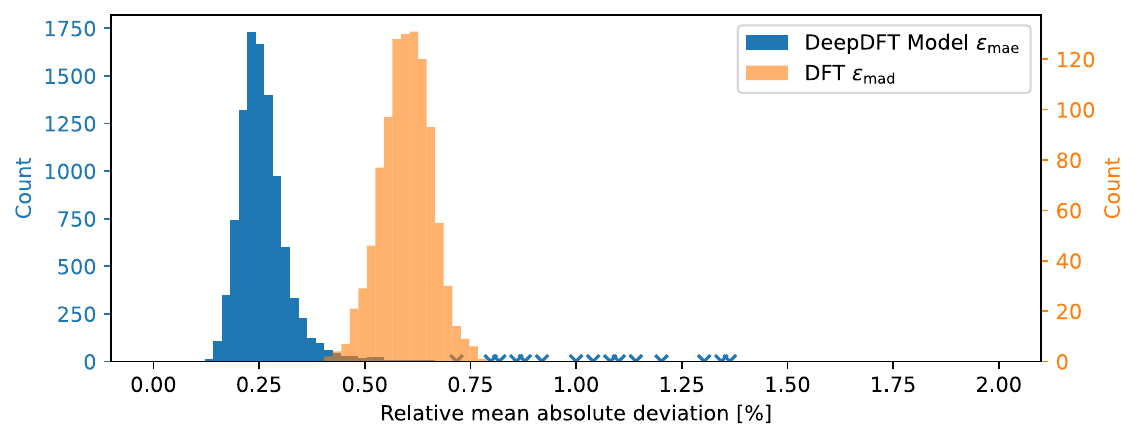}
	\caption{Histogram of test errors on QM9 in comparison with variations in DFT computed density using different exchange correlations functional. The markers show bins with a count of one. }
	\label{fig:qm9_test_errors_histogram}
\end{figure}

\begin{figure}[tbp]
	\begin{subfigure}[b]{0.49\textwidth}
		\includegraphics[width=\textwidth]{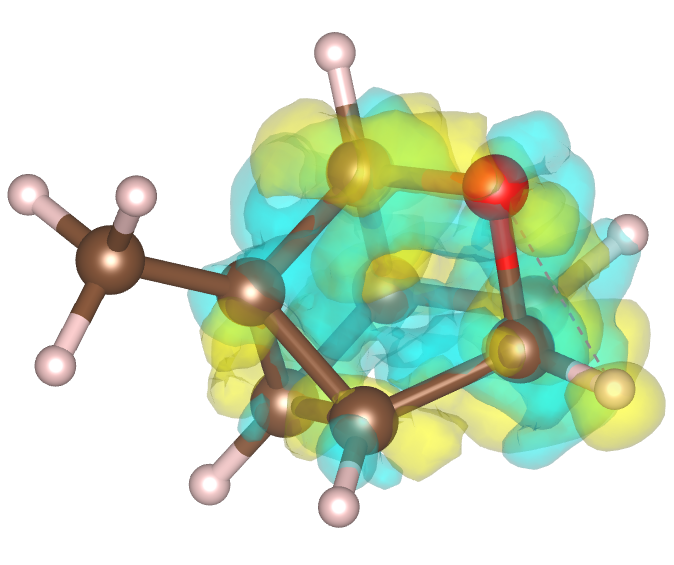}
		\caption{\ce{C8H8O} (1.1\%)}
		\label{fig:highest_error}
	\end{subfigure}
	\begin{subfigure}[b]{0.49\textwidth}
		\includegraphics[width=\textwidth]{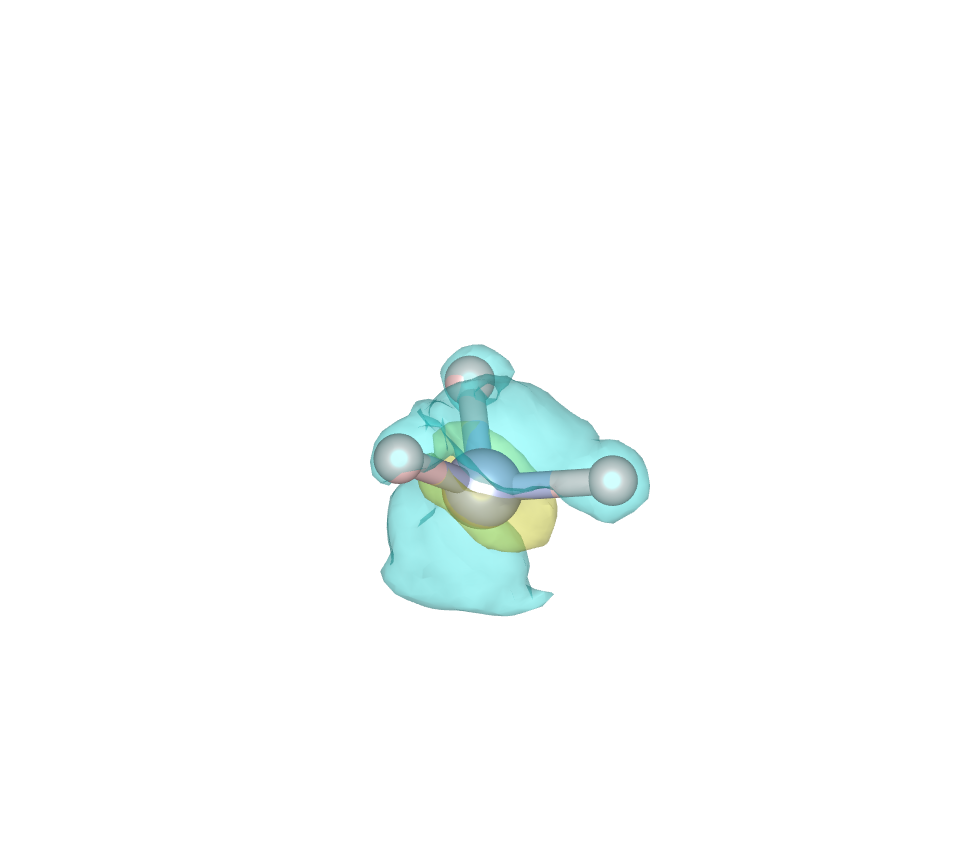}
		\caption{\ce{NH3} (1.0\%)}
		\label{fig:high_error}
	\end{subfigure}
	\begin{subfigure}[b]{0.49\textwidth}
		\includegraphics[width=\textwidth]{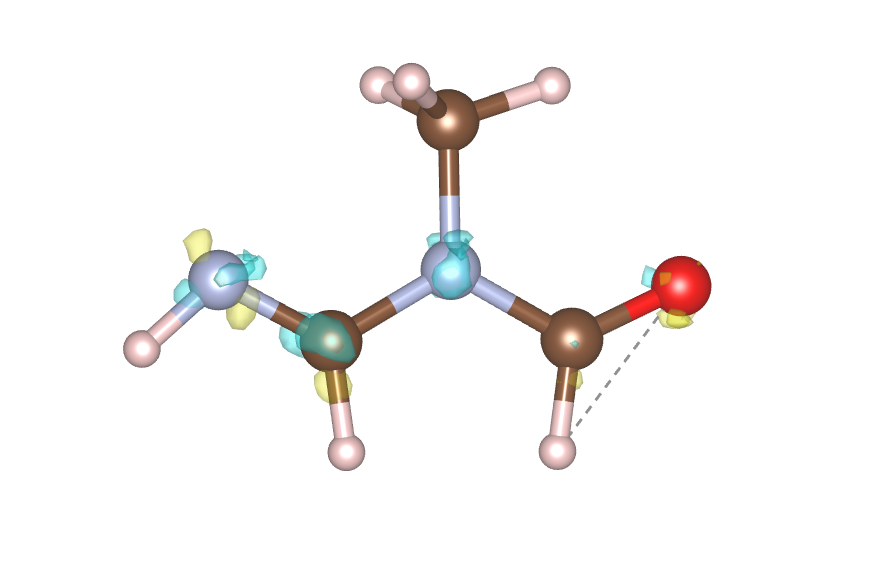}
		\caption{\ce{C3H6N2O} (0.28\%)}
		\label{fig:avg_error}
	\end{subfigure}
	\begin{subfigure}[b]{0.49\textwidth}
		\includegraphics[width=\textwidth]{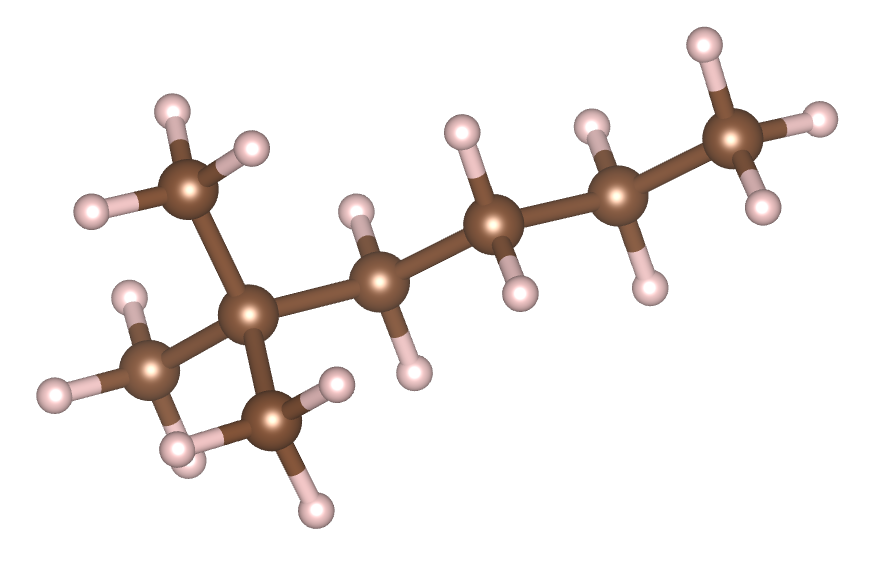}
		\caption{\ce{C8H18} (0.13\%)}
		\label{fig:low_error}
	\end{subfigure}
	\caption{Prediction error isosurfaces \SI{\pm 0.001}{e~Bohr^{-3}} for four example QM9 molecules from the test set with hydrogen (white), carbon (brown), oxygen (red) and nitrogen (blue). The error percentages in parenthesis denote the normalized mean absolute error $\varepsilon_{\mathrm{mae}}$ for each molecule. The chosen examples are two very high error molecules (\subref{fig:highest_error}) and (\subref{fig:high_error}), an example with average error (\subref{fig:avg_error}) and the molecule with the lowest error of all in the test set (\subref{fig:low_error}).}
	\label{fig:error_isosurfaces}
\end{figure}

The equivariant DeepDFT achieves notably lower prediction error than the invariant model, especially for the ethylene carbonate dataset. %
Both methods produce roughly two orders of magnitude lower error than the superposition of atomic densities baseline
. The QM9 accuracy is below that of OrbNet-Equi\cite{qiaoUNiTEUnitaryNbody2021}, which uses additional input information, i.e. the results of GFN1-xTB semiempirical electronic structure method are used as input features. Thus it can not be considered as a pure data driven model.
As %
another
 reference, we also look at the variation in electron density between DFT simulations performed using different exchange-correlation functionals in VASP. This can be seen as an appraisal of variability in DFT derived charge density. We randomly select 1000 examples of the QM9 test set and recalculate them with eight different XC functionals in VASP (BEEF, Perdew-Burke-Ernzerhof, Perdew - Wang 91, Ceperley-Alder, Perdew-Zunger, revised Perdew-Burke-Ernzerhof, revPBE, PBEsol) \cite{wellendorff2012density, perdew1996generalized, perdew1992accurate, ceperley1980ground, perdew1981self, hammer1999improved, zhang1998comment, csonka2009assessing}.
For every grid point, we select the median across the eight calculated densities as the reference point and compute the mean absolute deviation around this point as shown in \eqref{eq:mad}.
\begin{equation}
	\varepsilon_|mad| = \frac{\int_{\vec{\brr} \in V} \frac{1}{K} \sum_{k=1}^K  \abs{ {\rho}_k (\vec{\brr}) - \rho_|median| (\vec{\brr})  }}{\int_{\vec{\brr} \in V} \abs{\rho_1(\vec{\brr}) }}
	\label{eq:mad}
\end{equation}
For every molecule, the deviation is numerically integrated over the simulated volume.
This serves as an estimate of inherent variations in DFT derived density and is analogous to the error measure \eqref{eq:mae} used for the machine learning models.
The DFT variation of 1000 molecules is shown along with the prediction test set (10000 molecules) errors in \figref{fig:qm9_test_errors_histogram}.
The average DFT variation across molecules is \SI{0.60}{\percent}, which is generally higher than the DeepDFT error.
However,
 notice that the distribution of errors from the DeepDFT model has a much longer tail than DFT variation. There is even a single test point with \SI{11}{\percent} error (not visible on the histogram figure).
This is not the case for the two other datasets and is caused by the large chemical and structural variations within the QM9 dataset.
Modeling the uncertainty and detecting outliers is therefore important for future models.
Error isosurfaces for four example molecules are shown in \figref{fig:error_isosurfaces}. We are showing two very high error examples (\figref{fig:highest_error} and \figref{fig:high_error}), an average error datapoint (\figref{fig:avg_error}) and the lowest error example (\figref{fig:low_error}) in the test set.
The highest error example (\figref{fig:highest_error}) is a clear outlier and has very unnatural bond angles.
The ammonia example (\figref{fig:high_error}) is an isolated group and is therefore not well represented in the training data.
The average and lowest error points (\figref{fig:avg_error} and \figref{fig:low_error}) are well represented in the dataset and especially the hydrocarbon (\figref{fig:low_error}) is easily learned. Previous work focusing only on hydrocarbons could create models reaching an accuracy of \SI{1.26}{\percent} using less than 1000 DFT densities\cite{kamalChargeDensityPrediction2020}. Our models are more accurate for hydrocarbons while generalizing to other molecules as well. 

\begin{figure}[tbp]
	\begin{subfigure}[b]{0.3\textwidth}
	\includegraphics[width=\textwidth]{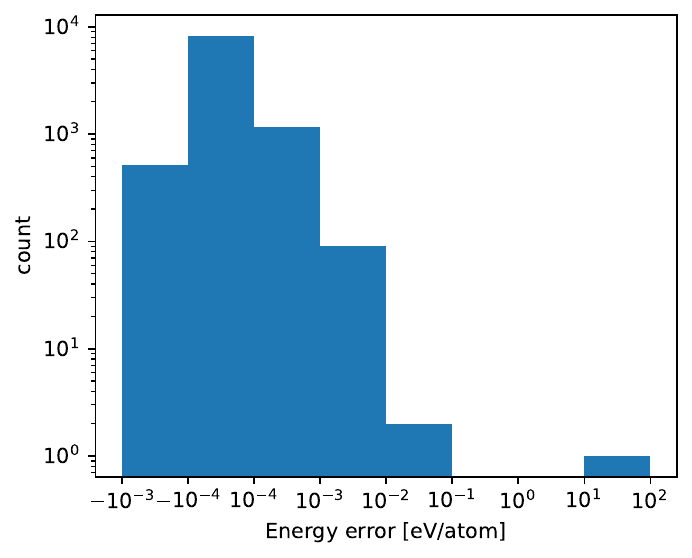}
	\caption{QM9}
	\label{fig:qm9_energy_errors}
	\end{subfigure}
	\begin{subfigure}[b]{0.3\textwidth}
	\includegraphics[width=\textwidth]{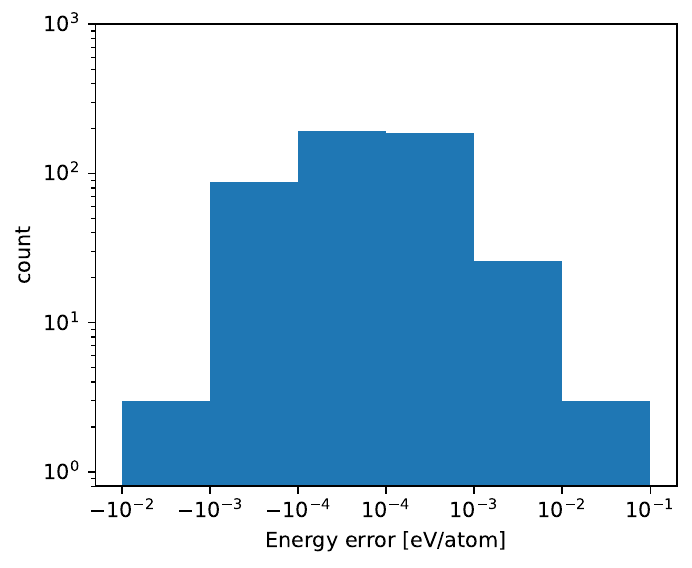}
	\caption{LIB Cathode}
	\label{fig:nmc_energy_errors}
	\end{subfigure}
	\begin{subfigure}[b]{0.3\textwidth}
	\includegraphics[width=\textwidth]{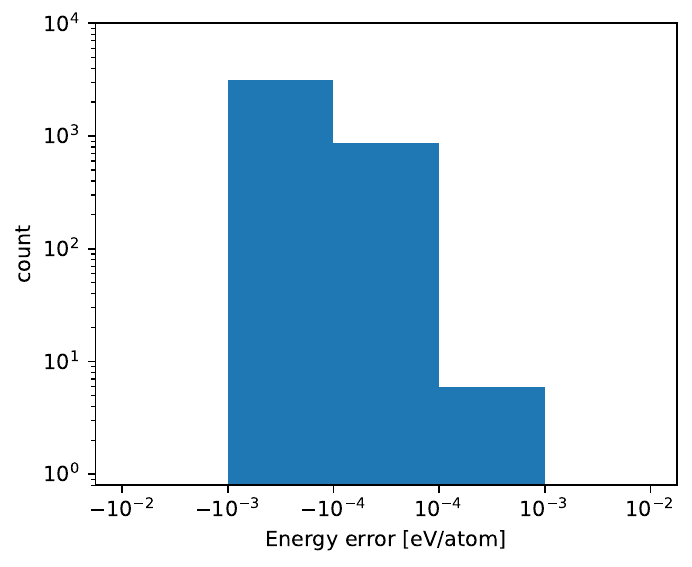}
	\caption{Ethylene Carbonate}
	\label{fig:ec_energy_errors}
	\end{subfigure}
	\caption{%
Distribution of energy errors obtained from running non-self-consistent single point energy calculations with VASP using the predicted DeepDFT charge densities. The distribution of normalized (by number of atoms) energy errors are shown for three different datasets QM9 (\subref{fig:qm9_energy_errors}), LIB Cathode (\subref{fig:nmc_energy_errors}) and Ethylene Carbonate (\subref{fig:ec_energy_errors}). The hisograms use logarithmic bins and count scale to clearly show both the heavy concentration around \SI{0}{eV/atoms} and the long, narrow tail of errors.
}
	\label{fig:energy_errors}
\end{figure}

To understand the accuracy of the model from an energy perspective, we take all the charge densities of the test sets and replace the density values with those predicted by the equivariant DeepDFT model (while keeping the PAW augmentation charges fixed), and run a single-point non-self-consistent energy calculation using VASP.
The obtained energies are compared with the self-consistent energies and the energy errors are shown in \figref{fig:energy_errors}.
The distribution of energy errors is heavily distributed around 0 with a long narrow tail, so we show the distribution in a log-log histogram.
For the QM9 dataset there are three molecules out of \num{10000} with absolute error above \SI{1e-2}{eV/atom}, while the mean absolute error for the remaining molecules is \SI{8.5e-5}{eV/atom}. For the two other datasets we observe less extreme outliers, but the tails are still present. The MAEs are \SI{4.2e-4}{eV/atom} and \SI{1.2e-4}{eV/atom} for the LIB cathode and ethylene carbonate datasets, respectively.

\begin{table}[tp]
  \caption{Datasets and prediction errors for invariant and equivariant DeepDFT model. %
We also compare with OrbNet-Equi and the superposition of atomic densities as implemented in VASP.
}
  \label{tab:datasets}
  \centering
	\begin{tabular}{lrrrSSSS}
\toprule
& \multicolumn{3}{c}{Dataset Splits} & \multicolumn{4}{c}{Test Set Error ($\varepsilon_|mae| \%$)}       \\
\cmidrule(r){2-4}
\cmidrule(r){5-8}
	Dataset  & {Train} & {Val.} & {Test} & {invDeepDFT} & {eqDeepDFT} & {OrbNet-Equi\cite{qiaoUNiTEUnitaryNbody2021}} & {Init VASP} \\
\midrule
	QM9 & 123835   & 50 & 10000     &  0.36  & 0.27& 0.21 & 15\\
	LIB Cath.    & 1450 & 50 & 500 &  0.09  &  0.06 & & 7.1\\
	Eth. Carb. &   7330    &  50 & 4000  & 0.53 & 0.18 & & 13 \\
\bottomrule
\end{tabular}

\end{table}

\subsection{Learning curve}
\begin{figure}[tp]
	\centering\includegraphics[scale=1.0]{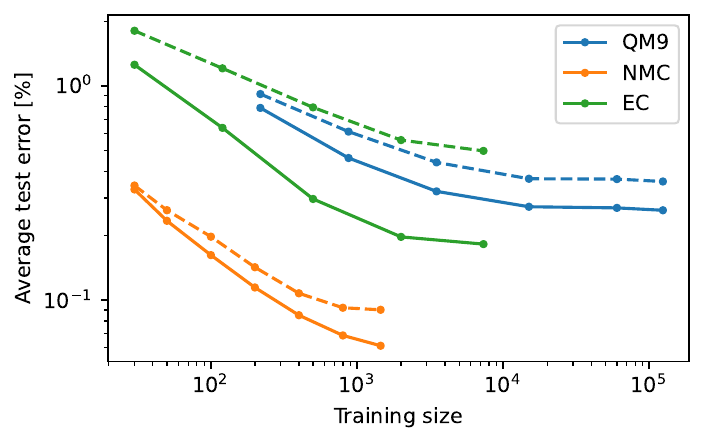}
	\caption{Learning curves for invariant (dashed lines) and equivariant (solid lines) models for three different datasets. The plot shows the average normalized test errors \eqref{eq:mae} for the QM9 (blue) LIB Cathode (orange) and Ethylene Carbonate (green) datasets as functions of the number of training examples, which are randomly sampled subsets of the full training sets.}
	\label{fig:learning_curve}
\end{figure}

In the sections above, we have looked at the average test errors for specific training and test set sizes.
To better understand the effectiveness of the learning method and the data efficiency of the models, it can be very useful to look at learning curves \cite{Huang2018}, i.e. test error as a function of the training set size plotted on a log-log scale.
The validation and test sets are the same as above, but we randomly sample a subset of the training data to reduce the training set size.
The result of this numerical experiment is shown in \figref{fig:learning_curve}.
Ideally the learning curves should follow a straight line in the log-log plot and be as steep as possible \cite{Huang2018}.
Initially all the learning curves are steep, but they flatten out with an increasing number of training examples.
The flattening of a learning curve is usually caused by either noise in the data, by a non-unique input representation or by lack of flexibility in the model.

If the saturation was caused by noise we would expect both models to converge to the same error. In contrast, for all three datasets, the equivariant models outperform the invariant models.
The input information presented to the equivariant models is the relative positions of atoms and their atomic numbers,
which is (given a large enough cutoff) enough to distinguish between all the inputs.
This indicates that the model lacks the flexibility to capture all the details and inter-dependencies modeled by density functional theory and this is not surprising, given the simpler architecture of the deep learning models and the approximate nature of the learned functions.
The training curves show (%
see Supplementary Figure 1 and Supplementary Figure 2
) that with \num{15000} training examples or more the model is not able to significantly overfit the QM9 dataset, but in all other cases the root mean squared error (RMSE) is lower on the training set than on the validation set.
During development we have briefly tried to double the number of interaction layers in the equivariant model and increased the cutoff radius to \SI{5}{\angstrom}, but we did not see an improvement in the accuracy of the model.
However, it might be possible to improve the model with a more systematic hyperparameter search or by introducing higher order equivariant internal representations as used in recently developed interatomic potentials, for example NequIP \cite{batznerEquivariantGraphNeural2022}.

Improved models are expected to provide higher accuracy and our analysis encourages further research and development of even more expressive deep learning architectures.

Among the three datasets, the ethylene carbonate dataset shows the largest difference in error between the two models. The added representation of directionality in the equivariant model is better at modeling highly polar molecules and the intermolecular interaction between them, which is further discussed in the following section.

\subsection{Intermolecular interactions}
\begin{figure}[htbp]
	\includegraphics[width=0.49\textwidth]{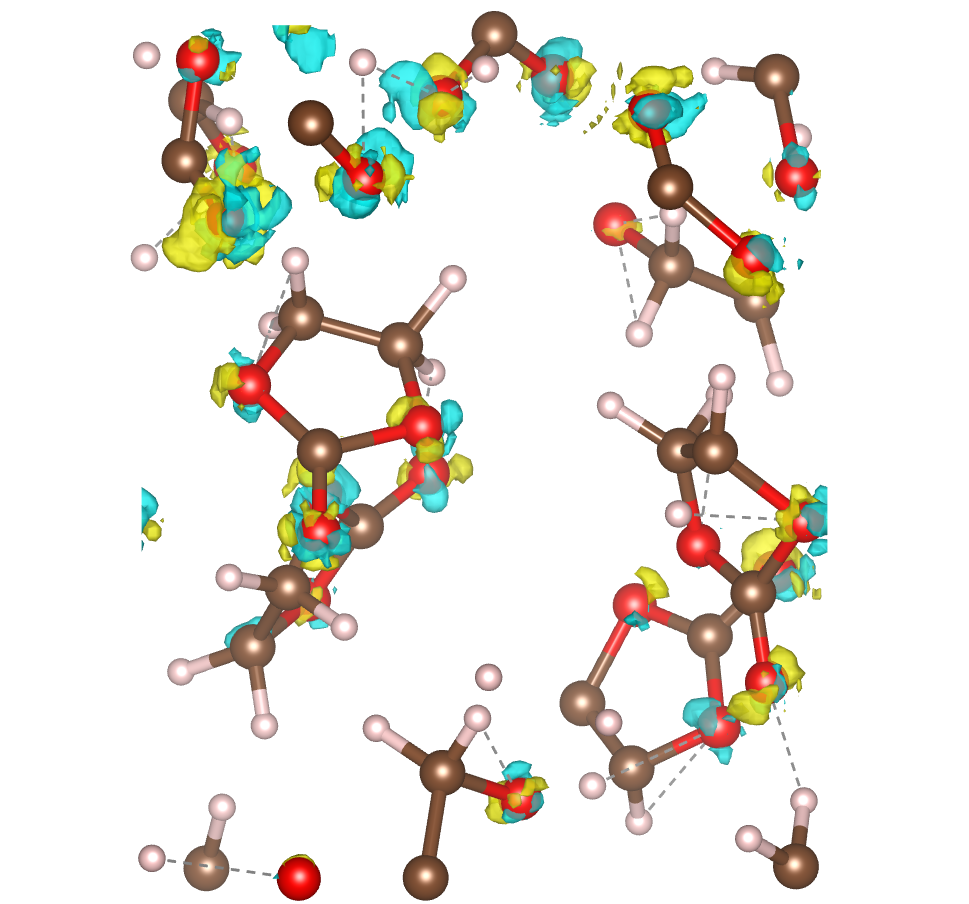} \hfill
	\includegraphics[width=0.49\textwidth]{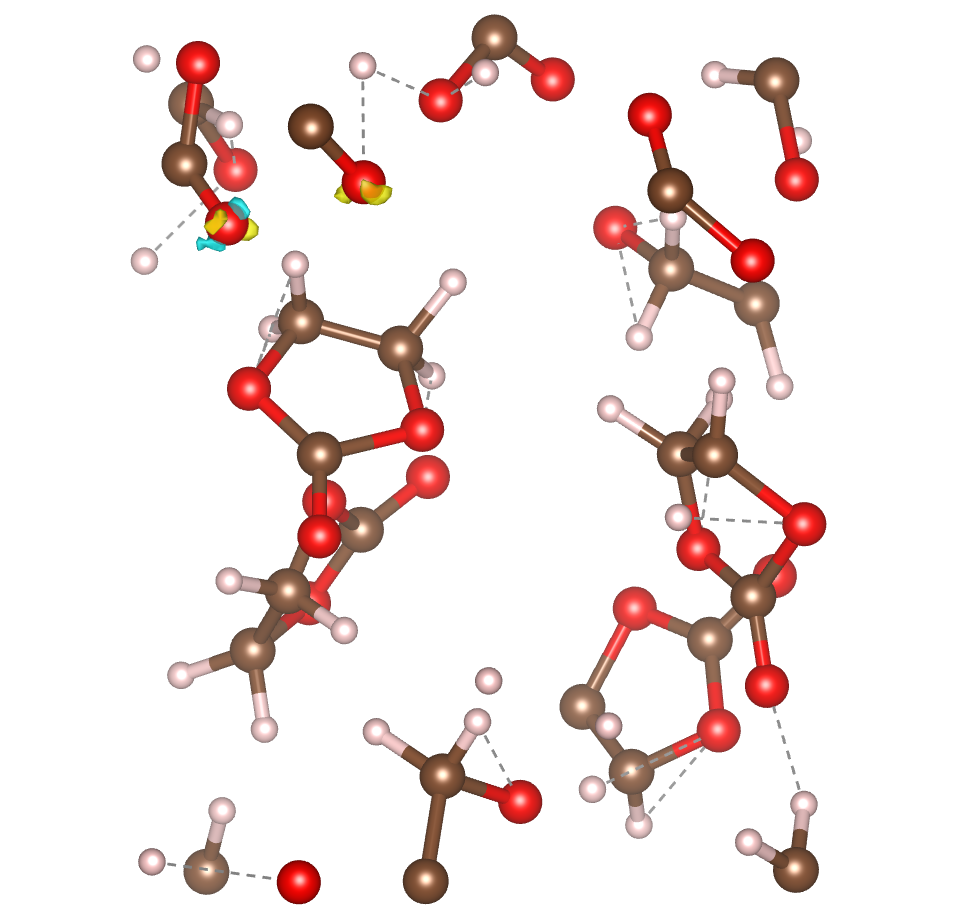}
	\caption{Prediction error isosurfaces of \SI{\pm 0.003}{e~Bohr^{-3}} with model based on invariant representation to the left and model based on equivariant representation to the right. The ethylene carbonate molecule contains hydrogen (white), carbon (brown) and oxygen (red).}
	\label{fig:ethyelenecarbonate_isosurfaces}
\end{figure}

To visualize the difference in prediction accuracy between the two models trained on the liquid ethylene carbonate structures, in \figref{fig:ethyelenecarbonate_isosurfaces} we show an error isosurface for one of the test set examples.
We see that most of the error is located around the oxygen atoms (red atoms).
To investigate this quantitatively we partition the electron density into volumes around each atom according to Bader partitioning\cite{tang2009grid}.
For the 4000 test examples this leads to 29\%, 16\%, 55\% of the total volume and 12\%, 20\%, 68\% of the total electron charge to be assigned to H, C, O respectively.
To understand how different elements contribute to the overall error, we calculate $\varepsilon_|mae|$ following \eqref{eq:mae}; but for each atom type we only integrate over the Bader volumes associated with atoms of that type.
Instead of normalizing with respect to the target density for each atom type in \eqref{eq:mae}, we also normalize with respect to the total error and calculate the total error share. The error decomposition for the invariant and equivariant models are shown in table \ref{tab:ec_error_decomp}.
\begin{table}[tp]
	\caption{Prediction errors of liquid ethylene carbonate data test set decomposed into Bader volumes for each atom type.}
  \label{tab:ec_error_decomp}
  \centering
	  \begin{tabular}{lcrr}
    \toprule
	Model  & Element & $\varepsilon_|mae|$ & Total Error share \\ %
    \midrule
			& H & 0.71\% & 17\% \\%
Invariant 	& C & 0.41\% & 16\% \\%
			& O & 0.49\% & 66\% \\%
    \midrule
				& H & 0.32\% & 21\% \\%
Equivariant 	& C & 0.18\% & 20\% \\%
				& O & 0.16\% & 59\% \\%
    \bottomrule
  \end{tabular}

\end{table}

As the error isosurface figure also showed, the majority of the error is assigned to the oxygen atoms (66\% for the invariant model and 59\% for the equivariant model), but most of the target electron density is also found within the oxygen Bader volumes (68\%).
We also notice that out of the three elements it is the oxygen volume that benefits the most from using the equivariant model.

\subsection{Runtime and Scalability}
\begin{figure}[tp]
	\centering\includegraphics[scale=1.0]{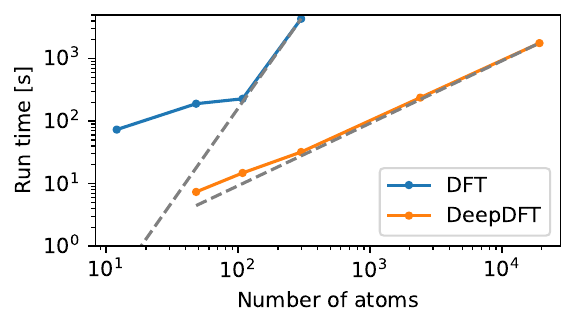}
	\caption{Computation time for VASP running on 2*20 core Intel Skylake Xeon CPU vs DeepDFT running on a single RTX 3090 GPU. The dashed lines show the expected asymptotic behaviours $ax^3$ and $ax$ respectively.}
	\label{fig:scalability}
\end{figure}

To demonstrate the scalability of the model we measure the runtime for calculating the electron density of systems of increasing sizes.
We use a single cell of 12 atoms \ce{Li3Co2NiO6} with periodic boundary conditions and repeat the unit cell to show how model run time and DFT run time scales with system size.
The result of the scalability test is plotted in \figref{fig:scalability}.
As expected we observe a linear trend for DeepDFT and it is therefore much faster than DFT for large systems, because of the cubic complexity of DFT. However, DeepDFT is also an order of magnitude faster even for small systems. Notice that DeepDFT is only running on a single GPU core and can be optimized to utilise more GPU cores in parallel for actual deployment towards high throughput tasks. The model is implemented in PyTorch for research purposes and even though it is orders of magnitude faster than density functional theory for large systems, it can still be made to run faster, e.g. utilising several GPUs in parallel when making predictions or by simplifying the readout network, because this part of the network needs to be run for point in the simulation grid and thus are completely parallelizable with no communication overhead. 

\subsection{Charge transfer in NMC Cathode}
Charge transfer in intercalation cathodes is not only a key descriptor for the ion intercalation process \cite{shen2020charge} and related energetics, but can also be used as a tool towards capturing degradation processes like oxygen evolution \cite{jung2017oxygen}. Thus understanding the electrochemical properties of cathode materials and computational screening for materials need access to charge density at various lithiated states.
In this section we investigate the applicability of the model to accurately track charge transfer in NMC cathode materials. To test the accuracy of the trained DeepDFT model, we randomly sample 30 structures from the NMC dataset (that are not part of the training dataset), remove one Li atom and relax the structure with DFT.
The DeepDFT model is then used to predict the electron density of the initial structure and of the one with one Li removed. If DeepDFT can capture the electron transfer redox process with good accuracy, we can use such models for screening optimal NMC compositions.
We do note that this numerical experiment is only meaningful for assessing the ability of the model to track charge transfer. In a real computational screening setting the structure relaxation would need to also be replaced by a machine learning model.

\begin{figure}[htbp]
	\centering\includegraphics[width=0.49\textwidth]{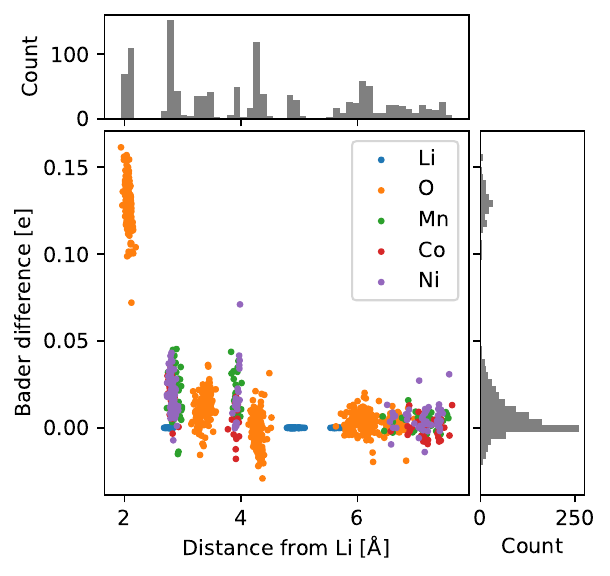}
	\hfill
	\centering\includegraphics[width=0.49\textwidth]{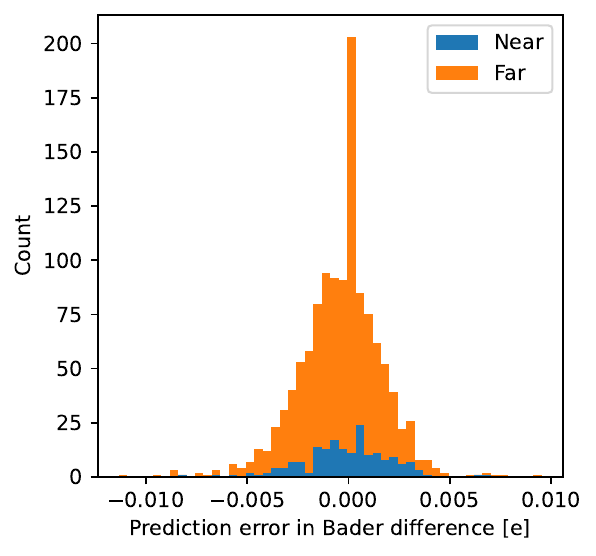}
	\caption{Analysis of Bader charge difference of remaining atoms when removing a Li atom from NMC cathode. The scatter plot to the left shows the change in Bader charge for all atoms in the systems as a function of the distance from the Li being removed. The stacked histogram to the right shows the distribution of prediction errors of the machine learning model.}
	\label{fig:bader_diff}
\end{figure}

To assign charge to each atom we use the Bader charge of each atom calculated with the Bader program\cite{baderprogram}. The change in charge of each atom for all the 30 systems is shown as a histogram in \figref{fig:bader_diff}.
Notice that we have two clusters of charge differences in the dataset.
The largest cluster is atoms that are not or very little affected by the removal of the Li atom.
The smaller cluster is the one that is influenced by the Li removal to a larger degree.
The prediction error is small for both clusters as shown in \figref{fig:bader_diff}.
This reinforces the capability of DeepDFT as a practical ML model that can be deployed for inexpensive large phase space exploration for high performance materials.
Instead of looking at individual errors we can also compute the total error of each structure. We calculate the sum of absolute errors in the charge difference across the unit cell and average across the 30 structures of the test set:
\begin{equation}
	\varepsilon_|total| = \frac{1}{30} \sum_{i=1}^{30} \sum_{n=1}^{\lvert Q^i\rvert} \| \Delta q^i_n - \Delta \hat{q}^i_n \|
	\label{eq:total_error}
\end{equation}
where $i$ is the index for the 30 systems in the test set, $\Delta q_n^i$ is the change in Bader charge of the $n$th atom of that system and $\lvert Q^i \rvert$ is the number of atoms in the $i$th system.
The average total error is \SI{0.060}{e}, which means that on average \SI{6.0}{\%} of the electron charge is distributed incorrectly. However, a large proportion of that error is due to fluctuations in charges far away from the removed Li. If we only consider the near atoms in the inner sum of \eqref{eq:total_error}, the error decreases to \SI{1.0}{\%}.

\section{Discussion}
In this work we have presented the equivariant DeepDFT model which achieves at par or beyond the state of the art  prediction accuracy on three widely different datasets including solid, liquid, and gas phases. Implementation of the equivariant framework (improvement on the invariant one \cite{jorgensen2020deepdft}) greatly helped in achieving high accuracy, especially in liquid state simulations as it helped learning better representations of inter- and intramolecular structure variations. 
Although the training is done on discrete grid points, the %
 learnt function is fully differentiable and thus can be used for further mathematical derivations, such as %
derivative-based visualization of interactions
using Interaction Region Indicator (IRI) \cite{iri2021} or
 Density Overlap Regions Indicator (DORI) \cite{dori}.  Additionally, the inference mechanism itself is parallelizable and can be distributed to a larger number of GPUs, which makes electron density simulation of millions of atoms feasible.

Surrogate models for energy and forces have helped perform high accuracy molecular simulations at long length and time scale or conduct high throughput screening at an unprecedented speed in the last few years. For functional materials discovery (e.g. battery electrodes, catalysts etc.) combining energy models with DeepDFT will let us model and improve properties where electron transfer redox reactions and charge density are critical for better functionality. %
Because of the low runtime and linear scaling with system size
 DeepDFT will let us explore far larger materials phase space than it is currently possible with DFT derived charge densities.

We foresee high throughput screening and large scale simulations pertaining to redox reactions and charge transfer properties will be tackled with DeepDFT in the near future. 

The error distribution for structures not well represented in the training data creates a long tail with high errors (as seen for a few unique molecules in QM9 dataset).
For using DeepDFT in high throughput studies and for building trained models for new classes of materials in a data efficient manner with active learning, it would be necessary to also model the uncertainty of the predictions, e.g. by using an ensemble of models.
In future it will also be beneficial to benchmark the model on other density derived properties such as the total energy, as utilizing density as the underlying variable can be more data efficient strategy than predicting the properties directly. The DeepDFT code is made available for the community and we are looking forward to seeing applications of the model in the simulation of materials and molecules as well %
as
 adaption into other models expanding on our codebase. 

\section{Methods}
\subsection{Equivariant Neural Message Passing Network}
\begin{figure}[htbp]
	\centering\includegraphics[width=0.5\textwidth]{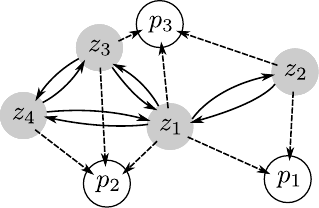}
	\caption{Neural message passing with probe nodes. Neural network computed messages are exchanged between atom vertices in several steps while the probe nodes only receive messages. The contents of the messages depend on the hidden state of the vertices and the startes are updated after each message-passing step.}
	\label{fig:msgpassing_model}
\end{figure}

In this section we describe more formally and in more detail how neural message passing is used to model the electron density around atoms.
The DeepDFT density model is framed in the neural message passing framework devised by \cite{gilmerNeuralMessagePassing2017}.
A simple example graph with four atoms and three query points is illustrated in \figref{fig:msgpassing_model}.
Each vertex has a hidden state that is updated in a number of interaction steps.

The previously introduced DeepDFT model \cite{jorgensen2020deepdft} uses an atom-to-atom interaction architecture very similar to the SchNet model \cite{schuttSchNetDeepLearning2018}.
However, in this type of model the hidden states of the vertices are scalar arrays, which contain no explicit information about directionality and are invariant to rotations. This for example, means that a change in angles can not be resolved through message passing steps \cite{pmlr-v139-schutt21a}.
This shortcoming has already been addressed by graph neural networks using spherical harmonics as the irreducible representation for the group of rotations in three dimensions \cite{thomas2018tensor, NEURIPS2019_03573b32, batznerEquivariantGraphNeural2022}. In these cases the hidden states become higher order tensors and they rotate (equivariantly) with the molecule.
A special case of equivariant neural networks, for which the equivariant states are Cartesian tensors, was recently introduced by \cite{pmlr-v139-schutt21a}.
In the so called polarizable atom interaction neural network (PaiNN) model the hidden vertex state contains a scalar array state as well as a vectorial state.
Equivariance of the vectorial state is conserved by restricting the vectorial states to interact only via cross products, inner products and scaling.
The
 new version of DeepDFT uses a variant of the PaiNN architecture as backend and the architectural details are given in this section. See the article introducing PaINN \cite{pmlr-v139-schutt21a} for more explanation on the method itself.

The input to the model are the atomic numbers $\left\{z_1, \cdots, z_N  \right\} \in \mathbb{N}$, the xyz-coordinate of each atom $\left\{\vec{\brr}_1, \cdots, \vec{\brr}_N\right\} \in \mathbb{R}^3$ and of each probe point $\left\{\vec{\bp}_1, \cdots, \vec{\bp}_M\right\} \in \mathbb{R}^3$ and in the case of crystal structure the unit cell vectors are also required $C \in \mathbb{R}^{3 \times 3}$. %
We use the arrow superscript $\vec{\bx}$ to emphasize vectors that are treated as geometric vectors, as opposed to arrays of scalars such as weight matrices.

From this information a directed graph is constructed with a vertex for each atom and another vertex type for each probe point.
The edges of the graph are drawn between atoms when they are within the cutoff distance (\SI{4}{\angstrom}) and incoming edges to the probe points are drawn when they are within the cutoff distance of an atom.
The atom scalar nodes are initialised with a learned embedding for each atom type
$\bss_i^0 = \ba_{z_i} \in \mathbb{R}^{F \times 1 }$ and the vectorial state is initialised to zeros $\vec{\bv}_i^0 = \vec{\bmm{0}} \in \mathbb{R}^{F \times 3}$.
The update in scalar state is given by a sum over messages from neighboring atoms
\begin{equation}
	\Delta \bss_i^m = \sum_{j \in N(i)} \bphi_s(\bss_j) \circ \bW_s( \vectornorm{\vec{\brr}_{ij}}) \cdot f_|cut|(\vectornorm{\vec{\brr}_{ij}})
	\label{eq:delta_s}
\end{equation}
where $\bphi_s$ is a 2-layer neural network with hidden layer and output layer size $F$, $\circ$ is element-wise vector multiplication and $\bW_s^m\left( \vectornorm{\vec{\brr}_{ij}} \right)$ is a continuous filter function. The feature-wise filter function is implemented as $F$ linear combinations of the distance expanded in sinc-like radial basis function \cite{klicpera_dimenet_2020} $\sin\left( \frac{n\pi}{r_|cut|}  \vectornorm{\vec{\brr}_{ij}} \right) / \vectornorm{\vec{\brr}_{ij}}$ with $1 \leq n \leq 20$. The cutoff function $f_|cut|(\vectornorm{\vec{\brr}_{ij}})= 0.5(\cos(\pi \vectornorm{\vec{\brr}_{ij}}/r_|cut|)+1) \textrm{ for } \vectornorm{\vec{\brr}_{ij}} < r_|cut| \textrm{ and } 0 \textrm{ otherwise}$ as proposed by \cite{behler_parinello}.
The cutoff function ensures a smooth transition when neighboring atoms enter the cutoff region.
The update in vectorial state is given by:
\begin{align}
	\Delta \vec{\bv}_i^m &= \sum_{j \in N(i)} \vec{\bv}_j \circ \bphi_|vv|(\bss_j) \circ \bW_|vv|\left( \vectornorm{\vec{\brr}_{ij}} \right) \cdot f_|cut|(\vectornorm{\vec{\brr}_{ij}}) \nonumber \\
	&+ \sum_{j \in N(i)} \bphi_|vs|(\bss_j) \circ \bW_|vs| \left( \vectornorm{\vec{\brr}_{ij}} \right) \frac{\vec{\brr}_{ij}}{\vectornorm{\vec{\brr}_{ij}}} \cdot f_|cut|(\vectornorm{\vec{\brr}_{ij}})
	\label{eq:delta_v}
\end{align}
The first sum is the convolution of scaled equivariant features $\vec{\bv}_j \circ \bphi_|vv|(\bss_j)$ with an invariant (only distance-dependent) filter function.
This is the only operation in the architecture where equivariant features are propagated through the network.
This allows directional information obtained through previous message passing interactions to propagate through the network.
The second term in \eqref{eq:delta_v} is the only operation in the architecture where new directional information is added to the hidden state.
Unit vectors corresponding to edges between vertices are scaled by $\bphi_|vs|(\bss_j)$ and by the invariant filter function $\bW_|vs| \left( \vectornorm{\vec{\brr}_{ij}} \right) f_|cut|(\vectornorm{\vec{\brr}_{ij}})$ and added to the vectorial node state.

For more expressiveness another two update equations are introduced, which operates atomwise across the scalar and vectorial features.
The updates for the scalar features are as follows
\begin{equation}
	\Delta \bss_i^{u} = \ba_|ss| (\bss_i, \vectornorm{\bV \vec{\bv}_i}) + \ba_|sv|\left( \bss_i , \vectornorm{\bV \vec{\bv}_i} \right)\left< \bU \vec{\bv}_i, \bV \vec{\bv}_i \right>
	\label{eq:delta_s_update}
\end{equation}
where $(\cdot, \cdot)$ means concatenation of features and $\ba_|ss|$ and $\ba_|sv|$ are 2-layer neural networks.
The vectorial features use the following update:
\begin{equation}
	\Delta \vec{\bv}_i^u = \ba_|vv|\left( \bss_i , \vectornorm{\bV \vec{\bv}_i}\right) \bU \vec{\bv}_i
	\label{eq:delta_v_update}
\end{equation}
which is a scaling of a linear combination of equivariant vector features.
The message passing algorithm works by computing \eqref{eq:delta_s} and \eqref{eq:delta_v} in parallel for all the node vectors and add the computed values to the current scalar and vectorial states respectively.
Then the nodes are updated by running \eqref{eq:delta_s_update} and \eqref{eq:delta_v_update} in parallel and update the nodes atomwise.
The message passing and update equations are repeated in several layers with different neural network weights at each layer.
In the PaiNN method and in this work we use 3 layers of message passing and update layers.

The hidden state of the special probe vertices are initialised with zeros (the scalar features are zero and the equivariant features are the zero vector).
They use the same message passing and update equations as for the atom vertices above, but the neural network weights are not shared between the two.
Furthermore, instead of using the residuals in \eqref{eq:delta_s} and \eqref{eq:delta_v} directly, we introduce a gating network that determines which features of the message sum to include and which part of the features to ignore.
\begin{align}
	\bss_i^{\mathrm{new}} = \bG_s(\Delta \bss_i^m) \circ \bss_i  + (\mathbf{1}-\bG_s(\Delta \bss_i^m)) \circ \Delta \bss_i^m  \label{eq:new_scalar_gated} \\
	\vec{\bv}_i^{\mathrm{new}} = \bG_v(\Delta \bss_i^m) \circ \vec{\bv}_i  + (\mathbf{1}-\bG_v(\Delta \bss_i^m)) \circ \Delta \vec{\bv}_i^m \label{eq:new_vector_gated}
\end{align}
The gating neural networks are two-layer neural networks with SiLU activation function for the hidden units and a sigmoid output activation function.
This allows the network to ignore parts of the incoming messages dependent on the total sum of messages.
After the final interaction steps the final state of each probe vertex is mapped to a single scalar for each probe state $\bss_i$. As in the original PaiNN model we also use a two layer neural network with SiLU activation function on the first layer and a linear activation function on the output layer. In principle the vectorial representation also enables prediction of vectorial properties at each probe vertex or higher order tensors constructed from a rank-1 tensor decomposition, as described in the original PaiNN article \cite{pmlr-v139-schutt21a}.

\subsection{Model parameters and training/validation setup}
In all experiments we use a feature size $F=128$ and cutoff distance $r_|cut|=\SI{4}{\angstrom}$.
When using the invariant message passing model we set $T=6$ and use $T=3$ for the equivariant message passing model.
The invariant model then has \num{2.1E6} parameters and the equivariant model \num{1.5E6} parameters.
Because of the large memory requirement for the electron density %
(CHGCAR)
 files, we use a rotating pool of 20 atomic configurations during training. New configurations are continuously loaded from the full dataset on disk into the rotating pool.
In each training step we sample two atomic structures and for each structure 1000 probe points are uniformly sampled from the %
VASP electron density grid
.
The cost function is the mean squared error of the probe points.
The Adam optimizer is used with initial learning rate $10^{-4}$ and the learning rate is exponentially decayed during training, i.e. the learning rate is $0.96^{s\cdot 10^{-5}}$ at gradient step number $s$. The validation set is used for early stopping. The cost on the validation set is computed every 5000 gradient steps and the final model is the one with the lowest error on the validation set. To reduce the variance and improve computational efficiency the probes of the validation set are kept fixed during the training and we use 5000 probes for each configuration.

\subsection{Datasets}
\begin{figure}[tbp]
	\begin{subfigure}[b]{0.31\textwidth}
		\includegraphics[width=\textwidth]{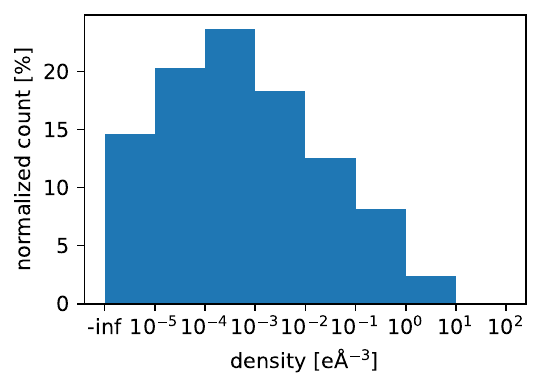}
		\caption{QM9}
		\label{fig:qm9_stats}
	\end{subfigure}
	\begin{subfigure}[b]{0.31\textwidth}
		\includegraphics[width=\textwidth]{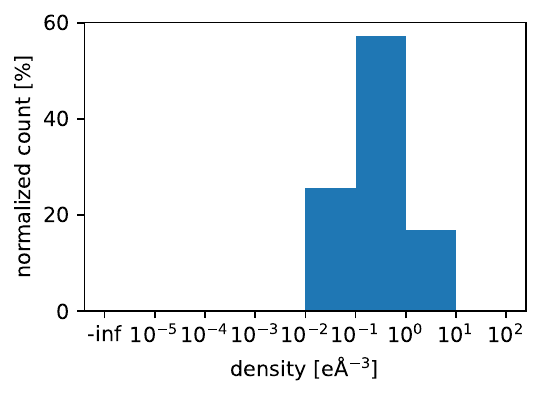}
		\caption{NMC}
		\label{fig:nmc_stats}
	\end{subfigure}
	\begin{subfigure}[b]{0.31\textwidth}
		\includegraphics[width=\textwidth]{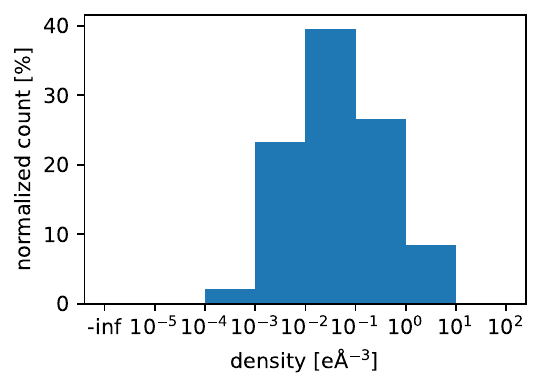}
		\caption{Ethylene carbonate}
		\label{fig:ec_stats}
	\end{subfigure}
	\caption{%
Normalized histograms with for the data examples of the three electron density datasets. QM9 (\subref{fig:qm9_states}) LIB Cathode (\subref{fig:nmc_stats}) and Ethylene Carbonate (\subref{ec_stats}). The histograms use logarithmic bins and show the number of data examples that fall into each bin as a percentage of the total number of examples.
}
	\label{fig:data_distribution}
\end{figure}

For all datasets the electron densities are calculated with VASP which is a projector-augmented wave based implementation of DFT. \SI{400}{eV} is used for the wavefunction cutoff and a Gaussian smearing of 0.1 eV is used for the electronic states. The first Brillouin zone is sampled only at the zone center for QM9 molecules, with a 3x3x1 Monkhorst–Pack k-point mesh for NMC data and 2x2x2 for liquid ethylene carbonate simulations. Perdew-Burke-Ernzerhof (PBE) exchange-correlation functional is used for all datasets. However to understand the effect of exchange correlation-functional on the variability of DFT calculated charge density we did DFT simulations with eight different functionals for small subset of QM9 molecules. Model training was done with PBE densities. 
The electron densities are output on a grid and the grid spacing is approximately \SI{0.1}{\angstrom} in all three datasets. The average number of grid points per configuration is 700k, 400k and 1M for the QM9, NMC and ethylene carbonate datasets, respectively. The distribution of the data points for each dataset is shown in \figref{fig:data_distribution}. The NMC dataset (\figref{fig:nmc_stats}) is the most dense dataset while the QM9 dataset (\figref{fig:qm9_stats}) has a lot of low electron density data examples. For datasets with even larger areas of ``empty space'' than the QM9 dataset it might be beneficial to remove some of the low density data points or in other ways weight the sampling of the training examples to avoid biasing the training towards predicting zeros.

\subsubsection{QM9 Dataset}
The geometries for the QM9 dataset \cite{Ramakrishnan2014-ey, Ruddigkeit2012-dc} are obtained from Figshare repository as deposited by \cite{ramakrishnan_dral_rupp_anatolevonlilienfeld_2014}. The VASP package only supports periodic boundary conditions so we use simulation cells with vacuum around molecules such that there is a gap of at least \SI{4}{\angstrom} or more.
As pointed out by a reviewer, the gap is not large enough to avoid all interactions between adjacent molecules. To quantify the interaction we have doubled the size of the unit cell of 1000 test set molecules and calculated the electron density at the same grid points as in the original dataset. The average $\varepsilon_|mae|$ (\eqref{eq:mae}) between the recalculated and original densities is $0.18\%$.

\subsubsection{NMC Dataset}
This dataset contains charge densities for NMC 2x2x1 supercell (12 transition metal atoms and 12 Li/vacancy site) with varying levels of Li content. For each structure we first randomly sample the number of Mn, Ni and Co atoms given that the total number of transition metal atoms is 12 and then randomly assign them to the transition metal positions of the lattice.
Similarly the number of vacancies is uniformly sampled between 0 and 12 and vacancies are assigned to the Li site. The generated configurations are then relaxed in two steps: First we relax the atom positions with fixed cell parameters and then we allow both positions and cell parameters to relax. We keep only the electron density (CHGCAR) file after the last cell relaxation step. The atoms are relaxed until forces on each atom are lower than \SI{0.01}{eV \per \angstrom}.

\subsubsection{Ethylene carbonate Molecular Dynamics Trajectory}
This dataset consists of charge densities of individual snapshots from a molecular dynamics trajectory. We insert 8 ethylene carbonate molecules in the simulation box.
To quickly explore a large part of the configurational space we put Hookean constraints on the molecular bonds(to maintain molecular identity such that molecules are not torn apart at such high temperature) and run Langevin molecular dynamics with thermostat temperature of \SI{3000}{\kelvin}. The simulation was run for 12380 steps of \SI{0.5}{fs} and we discard the first 1000 steps to reach equilibration.

\backmatter

\section{Data Availability}
The datasets are openly accessible at DTU figshare\cite{dataset_qm9vasp,dataset_nmc,dataset_ethylenecarbonate}.

\section{Code Availability}
A codebase for both models as well as pretrained PyTorch models are available on Github\cite{code_github}.

\section{Acknowledgments}
Peter Bjørn Jørgensen and Arghya Bhowmik acknowledges financial support from VILLUM FONDEN by a research grant (00023105) for the DeepDFT project.
\section{Author Contributions}
Arghya Bhowmik (A.B.) has designed and outlined the project and its research goal. Peter Bjørn Jørgensen (P.B.J.) and A.B. have designed and developed the model and method. P.B.J has carried out the software engineering, including implementation of model and training framework, implementation of data workflows for the datasets, and analysis of data and results. A.B. has supervised the research and designed the numerical experiments. A.B. and P.B.J. have both written the manuscript through mutual discussion.

\section{Competing interests}
The Authors declare no Competing Financial or Non-Financial Interests


%

%

\end{document}